\documentclass[twocolumns,reprint,letterpaper,amssymb,superscriptaddress,nobibnotes,showpacs, aps, prd]{revtex4-1}
\usepackage[dvipdfmx]{graphicx}
\usepackage{amsmath}
\usepackage{bm}

\begin{document}

\title{Search for kilogram-scale dark matter with precision displacement sensors}

\author{Akio Kawasaki}
\email{akiok@stanford.edu}
\affiliation{W. W. Hansen Experimental Physics Laboratory and Department of Physics, Stanford University, Stanford, California 94305, USA}

\begin{abstract}
The search for dark matter has been performed mainly for weakly interacting massive particles and massive compact halo objects, and the intermediate mass region has not been investigated experimentally. A method to search dark matter with precision displacement sensors is suggested for this mass range. The search is performed by detecting a characteristic motion of a test mass when it is attracted by a dark matter particle through gravity. Two different types of displacement sensors are examined: optically levitated microspheres and laser interferometers for gravitational wave detection. The state-of-the-art detectors' sensitivity is several orders of magnitude lower to put constraints on dark matter particles. Among the two types of detectors, gravitational wave detectors have higher sensitivities, and a sensitivity 10 times more than the next generation detector can potentially address the existence of dark matter particles of a few kilograms.
\end{abstract}

%\pacs{XX, XX, XX}
\maketitle

\section{Introduction}
There have been multiple independent astronomical observations that established the existence of dark matter (DM) \cite{AstrophysJ.238.471,%the rotational velocity of stars in a galaxy 
1807.06205,%cosmic microwave background
AstrophysJLett.648.L109}%gravitational lensing
, and different candidates of DM have been searched intensively: massive halo objects (MACHOs) \cite{AstrophysJLett.499.L9} and weakly interacting massive particles (WIMPs) \cite{1805.12562,PhysRevLett.118.021303,PhysRevLett.119.181302,PhysRevD.97.022002,EurPhysJC.76.25}, as well as even lighter particles such as axions and axionlike particles (ALPs) \cite{PhysRevLett.104.041301,PhysRevLett.112.091302}. So far, there has been no convincing discovery of the constituent of DM, and experiments and observations mainly set more and more stringent constraints on parameter spaces, as the sensitivity of detectors increases. 

Search methods are different according to the mass and the characteristics of DM candidates. MACHOs, which have a mass range around the solar mass ($\sim10^{30}$ kg), are searched by astronomical observations using gravitational lensing \cite{AstrophysJLett.499.L9}. The detection of WIMPs with the mass around $1-100$ GeV/c$^2$ ($10^{-27}-10^{-25}$ kg) is performed by detecting recoils of nuclei by the scattering with the DM particle \cite{1805.12562,PhysRevLett.118.021303,PhysRevLett.119.181302,PhysRevD.97.022002,EurPhysJC.76.25}. The mass of the axions and ALPs is typically assumed to be a few GeV or less, and these are typically detected by conversion from photons at accelerator beam dumps for relatively massive cases. Light mass ones ($\lesssim 1$ eV) are typically searched through the conversion to photons by a magnetic field. \cite{PhysRevLett.104.041301,PhysRevLett.112.091302,AnnRevNuclPartSci.65.485}. The intermediate mass scale, which is between $10^{-25}$ and $10^{30}$ kg, has not been intensively searched. This mass range includes various interesting particles and objects, such as the grand unification theory scale ($1.78\times 10^{-11}$ kg), Planck mass ($2.2 \times 10^{-8}$ kg), and primordial black holes ($10^{13}$-$10^{33}$ kg) \cite{PhysRevD.94.083504}. In this paper, a new method to search DM particles of this intermediate range is suggested. A precision displacement sensor works as a detector to observe the motion of a test mass, which can move by an attraction by DM particles, and two different kinds of displacement sensor are analyzed: optically levitated spheres, and gravitational wave detectors. 

\section{Displacement of a free test mass by a dark matter particle}
To think of how a test mass behaves when a DM particle interacts with it, we start from an analysis of a simple system consisting of a test mass and a DM particle. The DM particle, whose mass is $m$, is assumed to be a point particle or a particle of a size significantly smaller than its impact parameter $b$. The DM particle moves at a velocity of $v_0$ from aninfinitely distant place towards the test mass at rest, with an impact parameter $b$. The spherical test mass has mass $M$ and radius $r_0$, whose displacement is measured by a displacement sensor. The test mass is trapped around the origin by a harmonic trap of resonant frequency $\omega_0$ and damping constant $\gamma$. For simplicity, $r_0=0$, $\omega_0=0$, and $\gamma=0$ are assumed initially, and the effects of a harmonic trap and the finite size of the test mass are discussed later. The DM particle and the test mass interact only through the Newtonian gravity following $\mbox{\boldmath $F$}(t)=GMm{\mbox{\boldmath $r$}}(t)/r^3(t)$, where ${\mbox{\boldmath $r$}}(t)=\mbox{\boldmath $r$}_{\rm DM}(t)-\mbox{\boldmath $r$}_{\rm det}(t)$, $\mbox{\boldmath $r$}_{\rm DM}(t)$ and $\mbox{\boldmath $r$}_{\rm det}(t)$ are the position of the DM particle and the test mass, respectively, at time $t$, and $G$ is the gravitational constant.

The motion of two bodies interacting by a central force is well analyzed \cite{LandauMechanics}, and the analytical solution of the displacement $\mbox{\boldmath $r$}(t)=(x(t),y(t))$ at the center of mass frame is parametrized by $\xi$ in the following form:
\begin{eqnarray}
x &=& a\left( \epsilon-\cosh \xi\right) \label{EQmotionX} \nonumber\\
y &=& a\sqrt{\epsilon^2-1} \sinh \xi \\
t &=& \sqrt{\frac{\mu a^3}{\alpha}} \left( \epsilon \sinh \xi -\xi \right), \nonumber
\end{eqnarray}
where $\mu=mM/(m+M)$ is the reduced mass, $\alpha=GMm$, $a=\alpha/2E$, $\epsilon=\sqrt{1+2EL^2/\mu\alpha^2}$ is the eccentricity of the trajectory, $E=\mu v_0^2/2$ is the total energy of the system, and $L=\mu v_0 b$ is the total angular momentum of the system. 
%see p.65 of idea book #5 for this summary
Note that the $x$ and $y$ axes of the coordinate system are set in the plane of motion, and the directions of two axes are defined so as for the initial relative velocity to be $\mbox{\boldmath $v$}_0=v_0/\epsilon(1,\sqrt{\epsilon^2-1})$. To get the position of the test mass in the laboratory frame $\mbox{\boldmath $r$}_{\rm M}=(x_{\rm M},y_{\rm M})$, a Galilei transformation of $\mbox{\boldmath $r$}_{\rm M}=\mbox{\boldmath $r$}_{\rm CM}-(m/(M+m))\mbox{\boldmath $r$}$ is applied, where $\mbox{\boldmath $r$}_{\rm CM}$ is the position of the origin of the center of mass frame. The position of the test mass in the laboratory frame is therefore described as
\begin{eqnarray}
x_{\rm M} &=& \frac{m}{m+M}a\left(e^{\xi}-\epsilon-\frac{\xi}{\epsilon}\right), \nonumber\\
y_{\rm M} &=& -\frac{m}{m+M} \sqrt{\epsilon^2-1}a\frac{\xi}{\epsilon}, \label{MotionInLab} \\
t &=& \frac{a}{v_0} \left( \epsilon \sinh \xi - \xi\right). \nonumber
\end{eqnarray}
This satisfies the initial velocity of DM particle as $\mbox{\boldmath $v$}_0=v_0/\epsilon(1,\sqrt{\epsilon^2-1})$. The asymptotic behavior of $\mbox{\boldmath $v$}_{\rm M}=\partial \mbox{\boldmath $r$}_{\rm M}/\partial t$ at $t\rightarrow \pm \infty$ is
\begin{eqnarray}
\mbox{\boldmath $v$}_{-\infty}&=&\left(0,0\right)\\
\mbox{\boldmath $v$}_{\infty} &=&\left(\frac{2m}{M+m}\frac{v_0}{\epsilon},0\right),
\end{eqnarray}
which implies that the total momentum transfer $\Delta \mbox{\boldmath $p$}=(\Delta p_x, \Delta p_y)$ from the DM particle to the test mass at the end of the collision is only in the $x$ direction, and $\Delta p_x = 2mMv_0/(m+M)\epsilon$. In the case of $\epsilon \gg 1$, this is simplified to $\Delta p_x=2GMm/v_0 b$, which is the same quantity as the momentum transfer calculated by an assumption that $m\gg M$ and therefore the DM particle flies straight.

This motion can be detected in two different situations: (i) observing the relaxation of the test mass displacement in the harmonic trap after it receives a momentum kick of $\Delta p_x$ (damped oscillation measurement) and (ii) performing real-time detection while the test mass is being accelerated by the attraction from the DM particle (transient measurement). A damped oscillation measurement happens when a DM particle passes by the test mass so quickly that the collision process happens within the minimum time step $\Delta t_{\rm aq}$ of the displacement measurement of the test mass. This condition is $\Delta t_{\rm aq} > b/v_0$ within a factor of $O(1)$. The detection is performed through observing the damped oscillation of the test mass in the harmonic trap with the initial displacement of $y_{\rm M}=0$ and the velocity of $v_{Mx0}=\Delta p_x/M$. The criterion for this to be detected is that the maximum amplitude of this oscillation $A$ is larger than a minimum displacement $A_{\rm min}$ that can be detected on top of the noise of the detector:
\begin{equation}
A=\frac{C v_{Mx0}}{\omega_0}=\frac{2Gm}{bv_0\omega_0}\alpha \geq A_{\rm min},
\end{equation}
where $C$ is a numerical factor that is a function of $\omega_0$ and $\gamma$. This results in a condition for the collision parameter $b$ that the signal can be detected when $b<b_{\rm max}=2GMC/A_{\rm min}v_0\omega_0$.
The total volume $V$ that is scanned over by an observation for time $t_{\rm ob}$ is $V=\pi b_{\rm max}^2 v_0 t_{\rm ob}$.
When there are $N$ signals of DM particles during the observation period, the number density of the DM particle $n$ is
\begin{equation}
n=\frac{N}{V}=\frac{N}{4\pi t_{\rm ob}} \frac{A_{\rm min}^2 v_0 \omega_0^2}{G^2 m^2 C^2} \label{EqLimitDampedOsc}
\end{equation}
$N=3.69$ should be chosen to address the sensitivity with 95 \% confidence level when there is no signal. 

In the transient measurement, the motion of the test mass under the influence of a DM particle is monitored continuously. In the general case, it is difficult to predict analytically the motion of a test mass in a harmonic trap attracted by a DM particle. Here, for simplicity, we assume that the timescale of the oscillation and the damping is significantly longer than that of the force on the test mass by a DM particle, which is justified when we consider gravitational wave detectors. This simplifies the situation to the observation of Eq. (\ref{MotionInLab}). As $x_{\rm M}$ has a larger displacement than $y_{\rm M}$, $x_{\rm M}$ is used for this analysis. The sensitivity of the detector is typically described by a power spectral density as a function of the frequency of noises, and if the signal is larger than the noise at some frequency range, the signal can be detected. To estimate the sensitivity, the position of the test mass is Fourier transformed. This gives $1/f$ behavior, where $f$ is the frequency of the signal, and the largest $b$ that makes the signal of the test mass displacement cross the noise spectral density of the detector gives $b_{\rm max}$. With the same argument as the damped oscillation measurement, the sensitivity is described as 
\begin{equation}\label{EqLimitTransient}
n=\frac{N}{\pi v_0 t_{\rm ob} b_{\rm max}^2}
\end{equation}

\section{Application to actual systems}
The calculation so far has ignored certain aspects of actual detectors, such as the finite size of the test mass and the detector. To consider them, applications to two kinds of detector are discussed: optically levitated microspheres \cite{PhysRevLett.117.123604,PhysRevA.97.013842,PhysRevA.96.063841,NatPhys.7.527,ApplPhysLett.19.283,PhysRevLett.109.103603,NatCommun.4.2743,PhysRevA.94.053821,PhysRevA.93.053801,PhysRevLett.117.173602,JOptSocAmB.34.1421,ProcNatAcadSci.35.14180} and interferometers for gravitational wave detection \cite{PhysRevD.93.112004,ClassQuantGrav.32.024001,ClassQuantGrav.29.124007}. The calculation in the idealized system is modified according to the properties of the detectors, and the sensitivity to the DM particle density is numerically derived. 

For numerical calculations, the velocity of the DM particle is assumed to be $v_0=2.2\times 10^5$ m/s \cite{PhysRevLett.120.041102}, and the density of DM $\rho_{\rm DM}=0.39$ GeV c$^{-2}$ cm$^{-3}$ \cite{JCosmolAstropartPhys.2010.004,PDG}. The number density $n$ of DM particles of mass $m$ is therefore 
\begin{equation}
n=\frac{\rho_{\rm DM}}{m}= \frac{0.695\times 10^{-21}}{m~{\rm [kg]}}~{\rm[ m^{-3}]},
\end{equation}
with an assumption that all DM is made of particle of mass $m$. 

\subsection{Optically levitated microspheres}
Optically levitated microspheres are used for force sensors, and their method to measure the force is to convert the displacement of a sphere into the force by using its mass and resonant frequency of a harmonic trap. Thus, this system works as a precision displacement sensor of microspheres. The resonant frequency of the sphere ranges from a few hundred hertz \cite{PhysRevA.97.013842} to a few kilohertz \cite{PhysRevA.93.053801}, and to trap the sphere stably, displacement feedback with a bandwidth of an order of magnitude larger than the resonant frequency is applied. Here, for simplicity, $C = 1$ is assumed, which holds as far as the quality factor $Q$ of the resonance is much larger than 1. $C$ decreases as $Q$ decreases, but even when $Q=1$, $C\simeq 0.4$, which means even in the highly damped situation the difference in $C$ is at most a factor of a couple. In fact, the system in Ref. \cite{PhysRevA.97.013842} has $Q\sim1$ when the feedback cooling to reduce the noise is implemented. Also, it is possible to make a sequence such that for a certain amount of time the feedback cooling is turned on and off to alternate the high $Q$ for the measurement and low $Q$ for cooling to do the measurement in a high $Q$ environment without too much noise. Thus, assuming $C=1$ is plausible to have an estimate on sensitivity that can range orders of magnitude. The data acquisition of the position is performed at an order of 1-10 kHz. Suppose the DM particle passes at most 0.1 m away from the test mass, which is justified later. Because the timescale of the interaction between the test mass and the DM particle is significantly shorter than that of the feedback and the data acquisition, all the motion due to the attraction by a DM particle happens in a single bin of the data acquisition, and thus the detection mode is the damped oscillation measurement. 

The detection sensitivity $A_{\rm min}$ is determined by the noise level at the resonant frequency. This is $2\times10^{-10}$ m/$\sqrt{\rm Hz}$ for Ref. \cite{PhysRevA.97.013842} at $\omega_0=250$ Hz and $1\times10^{-9}$ m/$\sqrt{\rm Hz}$ for Ref. \cite{PhysRevA.93.053801} at $\omega_0=7300$ Hz, both of which are with feedback cooling (i.e., highly damped). In the case of Ref. \cite{PhysRevA.93.053801}, largest $\omega_0$ among three orthogonal axes is used to be conservative, whereas Ref. \cite{PhysRevA.97.013842} has more or less the same $\omega_0$ for all three axes. Also, there are two other factors limiting the sensitivity. One is the size of the microsphere, and the other is the size of the detector. When the collision parameter $b$ is smaller than the radius of the sphere $r_0$, the sphere cannot be regarded as a point mass any longer. This reduces the effective size of the sphere for considering the force by a DM particle to radius $b$ when the DM particle is closest to the sphere, resulting in a smaller amount of motion due to the DM particle. To be conservative, this effect is estimated to be the reduction of momentum transfer by a factor of $(b/r_0)^3$. Thus, for the sensitivity curve, a factor of $(b_{\rm max}/r_0)^6$ is multiplied at the region where $b_{\rm max}<r_0$. $r_0$ is 2.4 $\mu$m for Ref. \cite{PhysRevA.97.013842} and 150 nm for Ref. \cite{PhysRevA.93.053801}.

When $b_{\rm max}$ is large, the DM particle can attract something other than the microsphere, which potentially gives a fake signal. In the extreme case where $b_{\rm max}$ is significantly larger than the size of the laboratory, a DM particle passing far from the detector simply pulls the whole experimental system, and it is difficult to estimate exactly how the momentum kick onto the microsphere converts into an actual signal. To avoid confusion due to the attraction on the other components in the setup by the DM particle, it is assumed that when $b_{\rm max}$ is larger than the size of the detector $r_{\rm D}$, the sensitivity region is determined by $r_{\rm D}$, not $b_{\rm max}$, which results in 
\begin{equation}
n=\frac{N}{\pi v_0 t_{\rm ob} r_{\rm D}^2}
\end{equation}
for $b_ {\rm max}>r_{\rm D}$. For the numerical calculation, $r_{\rm D}=0.1$ m is assumed, as the size of vacuum chamber, inside of which only the last aspheric lens to tightly focus the trapping laser beam is located in an experiment in Ref. \cite{PhysRevA.97.013842}, is on the order of 0.1 m. 

\subsection{Laser interferometers for gravitational wave detection}
Laser interferometers for gravitational wave detection have had significant improvement in the past decades. They have two arms each of which has an optical cavity to enhance the effective path length. The mirrors for the cavities serve as test masses. If a DM particle interacts with only a single mirror or has a larger effect on one mirror than the other, the displacement of the mirror is recorded as a signal. Advanced LIGO \cite{PhysRevD.93.112004} and Advanced VIRGO \cite{ClassQuantGrav.32.024001} are currently in operation, and KAGRA \cite{ClassQuantGrav.29.124007} is under construction. These detectors have similar sensitivities, and in this analysis, Advanced LIGO is used as a representative. The arm is 4 km long, and the most sensitive frequency range is 10-1000 Hz, which means both the transient measurement and the damped oscillation measurement are possible. 

For the damped oscillation measurement, the highest resonant frequency is at 9 Hz \cite{ClassQuantumGrav.29.235004}, and the noise level at this resonant frequency sets $A_{\rm min}$ as $5 \times 10^{-17}$ m/$\sqrt{\rm Hz}$ \cite{PhysRevD.93.112004}. The fact that this resonance is one of the two undamped resonant modes in the mirror suspension system makes it suitable for the damped oscillation measurement. It should be noted that a specific data processing or analysis method might need to be developed for this DM search, and there might be a decent amount of background that can be difficult to distinguish a signal by DM particles from, as this frequency is at the low end of the frequency that is paid attention to by the gravitational wave observations. Based on their mirror size, $r_0\simeq 0.1 $ m. The detector size $r_{\rm D}$ is set as $r_{\rm D}=2$ km, which is half of the length of an interferometer arm. This is because if the impact parameter $b_{\rm max}$ is larger than this, the two circles of radius $b_{\rm max}$ centered on two cavity mirrors start to overlap, which results in a volume covered by two mirrors smaller than $V=2\pi b_{\rm max}^2 v_0 t$. Another justification is that when the DM particle passes a few kilometers away outside of the cavity, the force on the two mirrors becomes close, and the amount of signal is reduced. An additional factor to be considered in the case of a gravitational wave detector is that it is primarily for the detection of one-dimensional displacement, and therefore the sensitivity of the detector to the DM particle oscillates on a daily basis according to the relative angle $\theta$ between $v_0$ and the sensitive direction of the detector. The reduction in the amount of motion of the test mass is a factor of $\cos \theta$, and the time average of this is $\frac{1}{2\pi}\int^{2\pi}_{0} |\cos\theta| d \theta= 2/\pi$. Also, the fact that there are totally four mirrors in one interferometer needs to be taken into account. Each mirror can be regarded as a free test mass, and naively the enhancement by the four mirrors would be 4. However, two of the four mirrors are reasonably close to the input optics, compared to the mirrors at the other end of 4-km-long arm. Therefore, this factor should be 3, because two mirrors on the input side are so close to each other that the circles of radius $b_{\rm max}$ centered at the mirrors overlap with each other at large $b_{\rm max}$. The argument that each mirror is viewed as a test mass implies that the sensitivity becomes higher proportionally to the total number of interferometers, assuming that each interferometer is farther apart than $b_{\rm max}$.

The performance of the transient measurement is estimated by comparing the $f^{-1}$ curve and the sensitivity curve \cite{PhysRevD.93.112004}. The minimum signal curve that is tangent to the sensitivity curve is $3 \times 10^{-18}/f$, and therefore $b_{\rm max}$ is given by $b$ that induces this amount of signal. The transient measurement also has the limitation due to the detector size $r_{\rm D}$ and the test mass size $r_0$. 

To see whether future experiments have any benefits for the DM search, the Einstein Telescope (ET) \cite{ClassQuantGrav.27.194002} is also analyzed, though it is only for the transient measurement. As far as analyzing the effect on the single mirror, important parameters are the same as the LIGO case. The minimum $f^{-1}$ curve tangent to the sensitivity curve is assumed to be $1\times 10^{-19}/f$, based on the sensitivity curve and an arm length of 10 km. Note that the sensitivity curve is different between ET-B \cite{0810.0604} and ET-C \cite{ClassQuantGrav.27.015003}, and ET-C has a better sensitivity in low-frequency region, which can lower the minimum $f^{-1}$ curve by a factor of $\sim2$. It is assumed that $r_{\rm D}$ equals 5 km, because of the 10-km-long arm, and $r_0=0.1$ m is assumed, as the size or the mirror is on the order of 10 cm. 

\subsection{Current and future sensitivities}
\begin{figure}[tb]
	\begin{center}
 \includegraphics[width=1\columnwidth]{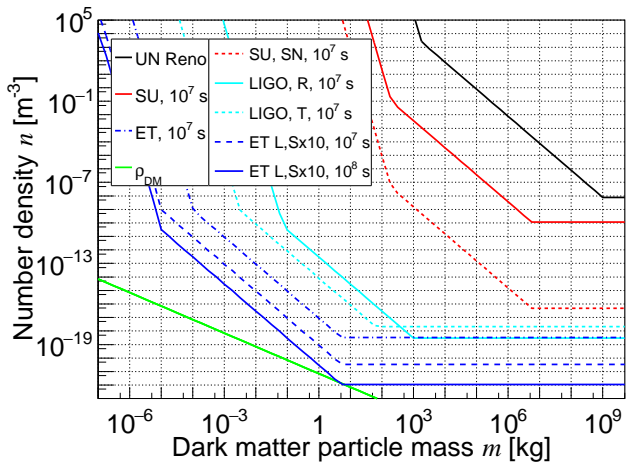}
 \caption{
Potential sensitivities of displacement sensors to dark matter particles of different mass without any background signals (95\% confidence level): UN Reno is the potential limit the data shown in Ref. \cite{PhysRevA.93.053801} can put. SU refers to the setup described in Ref. \cite{PhysRevA.97.013842} with $10^7$ s integration time by the current performance (red solid line) and by the improved noise level down to the shot noise limit (SN; red dotted line). The light blue solid line (LIGO, T, $10^7$ s) and light blue dotted line (LIGO, R, $10^7$ s) show the Advanced LIGO \cite{PhysRevD.93.112004} transient measurement and the resonant measurement with $10^7$ s integration time by a single mirror, respectively. Dark blue lines show the sensitivities for the Einstein Telescope \cite{ClassQuantGrav.27.194002}. The dashed dotted line (ET, $10^7$ s) is for single mirror, with $10^7$ s integration time. The dashed line (ET, L,S$\times 10$, $10^7$ s) and solid line (ET, L,S$\times 10$, $10^8$ s) are for 10 times higher sensitivity and detection radius than the current design for $10^7$ s with a single mirror and for $10^8$ s with three mirror sites, respectively. } 
 \label{FigLimit}
 \end{center}
\end{figure}

The sensitivity to the number density of DM particles with the 95 \% confidence level is summarized in Fig. \ref{FigLimit}. The black line (UN Reno) is the potential limit that can be set by currently available data in Ref. \cite{PhysRevA.93.053801}. The fact that their force measurement graph averages down over $10^5$ s proportionally to the inverse of the square root of the measurement time means that no extra feature in addition to the noise was observed, leading to the conclusion that there was no enormous signal. The actual limit has to be set by carefully analyzing the data to see if there are any small signals. 

Stanford curves are the expected performance based on Ref. \cite{PhysRevA.97.013842} with the measurement time of $10^7$ s. The solid line is the sensitivity with the current performance of the detector, and the dotted line assumes an improved position sensitivity limited by the shot noise. When the measurement time is the same as the UN Reno curve, the Stanford setup have 4 orders of magnitude higher sensitivity than the UN Reno setup as long as the sensitivity is not limited by the detector size. This is simply because the Stanford setup has better position sensitivity. This trend is the same when microsphere setups and gravitational wave detectors are compared. Advanced LIGO can put 9 orders of magnitude better constraint than the shot-noise-limited performance of the Stanford system, mainly because it has a smaller amount of noise at the mechanical resonance compared to the microsphere setups. Still, the sensitivity by the damped oscillation detection is 5 orders of magnitude above the number density of DM particles. 

As for the transient measurement, the sensitivity for a particle of the same mass is an order of magnitude higher than the resonant measurement, because the noise level relevant to the transient measurement is much better than that at the mechanical resonance at 9 Hz. Note that the transient measurement has to detect smaller motion than the overall amplitude at the resonant measurement, which prevents the sensitivity from being improved by the same amount of the noise ratio between two measurements. The sensitivity itself is still 5 orders of magnitude lower than the number density of the DM particles at $m=5$ kg. This is consistent with the analysis in Ref. \cite{PhysRevD.98.083019}. In their analysis, 10 and 1000 kg DM particles have a cumulative rate of $\sim10^{-5}$ yr$^{-1}$ and $\sim4\times10^{-4}$ yr$^{-1}$, respectively, at a signal-to-noise ratio (SNR) of 1. Figure 1 suggests that 3.69 hits are expected at a SNR of 1 over an observation of a third of a year, if the DM particle density were $10^6$ more than the estimate from the standard DM density. This is converted to $0.85\times10^{-5}$ yr$^{-1}$ cumulative rate. In the case of 1000 kg, if the sensitivity is not limited by the detector size $r_{\rm D}$ in Fig. 1, the attainable sensitivity is around $10^{-20}$ m$^{-3}$, which is 4 orders of magnitude larger than the actual DM particle density. With a similar calculation to the 10 kg case, this is equivalent to a cumulative rate of $8.5\times 10^{-4}$ yr$^{-1}$. Thus, the analysis shown here matches that in Ref. \cite{PhysRevD.98.083019} within a factor of $O(1)$. 

The ET is analyzed only for the transient measurement because of an unavailability of detailed information on the mechanical resonance. Thanks to the lower noise and longer arm, both the sensitivity at the same mass and the detector size limited sensitivity are improved compared to the LIGO case. However, it is still 3 orders of magnitude away from the DM particle density even at the closest point of 5 kg. When both the noise level is reduced by a factor of 10 and the detector size is increased by a factor of 10, the sensitivity improves by a factor of 100. With all three mirror sites taken into account and an observation performed for three years, the sensitivity can reach the DM particle density at $m=5$ kg. 

\subsection{Discriminating the signal from the background}
The discussion so far is simply based on the sensitivity limited by noises on the detector, but what limits the sensitivity is not only noises but also backgrounds that resemble signals. Because the detection of DM particles is performed by measuring small forces, a range of sources can induce background events. When a single test mass is used for the observation, moving objects on Earth, such as cars and airplanes, can induce background events. Particularly for the resonant measurement, it is extremely difficult to tell the signal from the background, as the momentum kick is assumed to happen instantaneously, and no information on the source of the momentum kick is recorded, except for the amount of the momentum kick. To sort such backgrounds out, taking coincidence between two or more test masses would help the discrimination of the background; a DM particle has a velocity of $v_0=2.2\times 10^5$ m/s that is significantly faster than a typical object moving on Earth (e.g., an airplane flies at the speed of less than the speed of sound, which is 340 m/s). In a simple case of an object moving along the line of two test masses, telling the DM particles from other backgrounds by velocity is easy. Even when the object is moving perpendicular to the line of two test masses, in which case the hits to these two test masses happen simultaneously, a third test mass at the position that makes an equilateral triangle together with the two other test masses would serve as a good source to tell the velocity of the object. As for LIGO, two test masses can be a pair of mirrors that makes a single arm, and the third test mass can be the mirror on the far end of the other arm of the interferometer. For the transient measurement, in principle, the time-dependent force should provide information on how fast an object is moving against the test mass even if there is a single test mass, but still timing information with two or more test masses should help to discriminate the signal from background, particularly when the signal is small.

Another source of background that cannot be avoided is the seismic noise and other kind of vibrations of continuous medium on Earth, such as air and water. There can be multiple ways of telling these apart from signals from a DM particle. One method is to have a seismometer or other kind of vibration meter. This can veto the detector when the vibration is beyond a threshold that is set independently. This method works even with a single test mass. When there is more than one test mass, discrimination by the velocity in the same way as a moving object is also possible. The propagation of these vibrations through these continuous media typically ranges between $10^2$ and $10^4$ m/s, which is orders of magnitude smaller than $v_0$.  

The coincidence of two independent backgrounds on two different test masses that mimics a single event by a DM particle still can happen, which is very difficult to tell apart from the signal. One way to reduce this kind of coincidence of backgrounds is to take a coincidence of as many test mass as possible. This is practically very difficult for gravitational wave detectors, as the existing detectors are significantly farther apart than their arm lengths. As for the optically levitated microspheres, locating several detectors in a desired way is possible, and in this case, extending $b_{\rm max}$ is important. Currently, to be conservative, $b_{\rm max}$ is assumed to be the distance between the sphere and the closest optics component, but potentially this can be made larger by fixing all optics components much tighter than the sphere in trap, i.e., making the resonant frequency of optics components much higher than the trapping frequency of the sphere. 

The noise on the detection system of the position of the test mass also should be considered. This would be mainly electrical transient backgrounds on detectors. The simplest way to reject this is to have an electrically independent system for each test mass. There might be some global glitches to all such independent systems, for example lightning and instantaneous power outage. These events should be able to be detected by some other methods, and the observation should be vetoed during these events.

\section{Implication to astrophysics and particle physics}
The DM particles in the mass range of a few kilograms have never been searched before. If these particles are point particles, or particles that are smaller than its Schwarzschild radius $r_{\rm s}=2GM/c^2$, these are black holes. Such small mass black holes are usually discussed in the context of the primordial black hole \cite{PhysRevD.94.083504}, but primordial black holes of kilogram scale mass evaporate quicker than the lifetime of the Universe, and therefore theoretically they have already been excluded. In case such small mass black holes can be generated at some time other than the birth of the Universe, this will be the first search for such light mass black holes. If the size of the DM particles is larger than $r_{\rm s}$, the DM can be an unknown particle that interacts with other matters only through gravity. 

An interesting mass range relatively close to the sensitivity plot is the Planck mass $ 2.18 \times 10^{-8}$ kg. At this mass, the sensitivity is 10 orders of magnitude above the number density of DM particles. Comparing the "ET, $10^7$ s" line and the "ET $L,S\times10$, $10^7$ s" line, reducing the noise level by a factor of 10 moves the sensitivity curve towards an order of magnitude smaller mass range. This implies that five orders of magnitude reduction of noise compared to the "ET $L,S\times10$, $10^8$ s" line is necessary to reach the Planck mass, which is unrealistic with the currently available technology. 

The discussion so far puts an emphasis on setting a constraint on the existence of DM particles when there is no signal including background and did not mention how the discovery of such a particle can happen. One method to narrow down the mass range is to detect the same DM particle by two test masses. In the general case, the impact parameters $b$ for the two test masses are different, and this gives the information on two unknown parameters $m$ and $b$. Careful analysis of gravitational wave detectors would easily give this information, as different cavity mirrors serve as different test masses, and having two or more optically trapped microspheres nearby would give similar information. In this sense, optically levitated microspheres have the advantage that it is easier to build multiple detectors aligned in a desired way, both for narrowing the mass range and for separating the signal from the background.

\section{Summary}
A method of using a precision displacement sensor of a test mass for a DM particle search is discussed. Although present and future technology by optically levitated microspheres can put a marginal constraint to the number density of DM particles of ton scale, future gravitational wave detectors that have a 10 times lower noise level and 10 times more detector size than the ET can potentially have high enough sensitivity to detect DM particles of $\sim 5$ kg. This would be a first experimental search for primordial black holes, and an even further improvement in the noise level by several orders of magnitude enables a search for DM particles of Planck mass. The optically levitated microspheres have lower sensitivities but have an advantage of easily making an array of test masses for the background discrimination.

\section*{aAcknowledgments}
The author thanks to Giorgio Gratta for an insightful discussion and acknowledges the partial support of a William M. and Jane D.Fairbank Postdoctoral Fellowship of Stanford University.

\bibliographystyle{apsrev4-1}
\bibliography{DMSearch}

\begin{thebibliography}{37}%
\makeatletter
\providecommand \@ifxundefined [1]{%
 \@ifx{#1\undefined}
}%
\providecommand \@ifnum [1]{%
 \ifnum #1\expandafter \@firstoftwo
 \else \expandafter \@secondoftwo
 \fi
}%
\providecommand \@ifx [1]{%
 \ifx #1\expandafter \@firstoftwo
 \else \expandafter \@secondoftwo
 \fi
}%
\providecommand \natexlab [1]{#1}%
\providecommand \enquote  [1]{``#1''}%
\providecommand \bibnamefont  [1]{#1}%
\providecommand \bibfnamefont [1]{#1}%
\providecommand \citenamefont [1]{#1}%
\providecommand \href@noop [0]{\@secondoftwo}%
\providecommand \href [0]{\begingroup \@sanitize@url \@href}%
\providecommand \@href[1]{\@@startlink{#1}\@@href}%
\providecommand \@@href[1]{\endgroup#1\@@endlink}%
\providecommand \@sanitize@url [0]{\catcode `\\12\catcode `\$12\catcode
  `\&12\catcode `\#12\catcode `\^12\catcode `\_12\catcode `\%12\relax}%
\providecommand \@@startlink[1]{}%
\providecommand \@@endlink[0]{}%
\providecommand \url  [0]{\begingroup\@sanitize@url \@url }%
\providecommand \@url [1]{\endgroup\@href {#1}{\urlprefix }}%
\providecommand \urlprefix  [0]{URL }%
\providecommand \Eprint [0]{\href }%
\providecommand \doibase [0]{http://dx.doi.org/}%
\providecommand \selectlanguage [0]{\@gobble}%
\providecommand \bibinfo  [0]{\@secondoftwo}%
\providecommand \bibfield  [0]{\@secondoftwo}%
\providecommand \translation [1]{[#1]}%
\providecommand \BibitemOpen [0]{}%
\providecommand \bibitemStop [0]{}%
\providecommand \bibitemNoStop [0]{.\EOS\space}%
\providecommand \EOS [0]{\spacefactor3000\relax}%
\providecommand \BibitemShut  [1]{\csname bibitem#1\endcsname}%
\let\auto@bib@innerbib\@empty
%</preamble>
\bibitem [{\citenamefont {Rubin}\ \emph {et~al.}(1980)\citenamefont {Rubin},
  \citenamefont {Ford},\ and\ \citenamefont {Thonnard}}]{AstrophysJ.238.471}%
  \BibitemOpen
  \bibfield  {author} {\bibinfo {author} {\bibfnamefont {V.~C.}\ \bibnamefont
  {Rubin}}, \bibinfo {author} {\bibfnamefont {W.~K.}\ \bibnamefont {Ford}}, \
  and\ \bibinfo {author} {\bibfnamefont {N.}~\bibnamefont {Thonnard}},\ }\href
  {http://adsbit.harvard.edu//full/1980ApJ...238..471R/0000471.000.html}
  {\bibfield  {journal} {\bibinfo  {journal} {Astrophys. J.}\ }\textbf
  {\bibinfo {volume} {238}},\ \bibinfo {pages} {471} (\bibinfo {year}
  {1980})}\BibitemShut {NoStop}%
\bibitem [{\citenamefont {Akrami}\ \emph {et~al.}(2018)\citenamefont {Akrami}
  \emph {et~al.}}]{1807.06205}%
  \BibitemOpen
  \bibfield  {author} {\bibinfo {author} {\bibfnamefont {Y.}~\bibnamefont
  {Akrami}} \emph {et~al.} (\bibinfo {collaboration} {Planck Collaboration}),\
  }\href {https://arxiv.org/abs/1807.06205} {\  (\bibinfo {year} {2018})},\
  \Eprint {http://arxiv.org/abs/arXiv:1807.06205} {arXiv:1807.06205}
  \BibitemShut {NoStop}%
\bibitem [{\citenamefont {Clowe}\ \emph {et~al.}(2006)\citenamefont {Clowe},
  \citenamefont {Bradač}, \citenamefont {Gonzalez}, \citenamefont
  {Markevitch}, \citenamefont {Randall}, \citenamefont {Jones},\ and\
  \citenamefont {Zaritsky}}]{AstrophysJLett.648.L109}%
  \BibitemOpen
  \bibfield  {author} {\bibinfo {author} {\bibfnamefont {D.}~\bibnamefont
  {Clowe}}, \bibinfo {author} {\bibfnamefont {M.}~\bibnamefont {Bradač}},
  \bibinfo {author} {\bibfnamefont {A.~H.}\ \bibnamefont {Gonzalez}}, \bibinfo
  {author} {\bibfnamefont {M.}~\bibnamefont {Markevitch}}, \bibinfo {author}
  {\bibfnamefont {S.~W.}\ \bibnamefont {Randall}}, \bibinfo {author}
  {\bibfnamefont {C.}~\bibnamefont {Jones}}, \ and\ \bibinfo {author}
  {\bibfnamefont {D.}~\bibnamefont {Zaritsky}},\ }\href
  {http://stacks.iop.org/1538-4357/648/i=2/a=L109} {\bibfield  {journal}
  {\bibinfo  {journal} {Astrophys. J. Lett.}\ }\textbf {\bibinfo {volume}
  {648}},\ \bibinfo {pages} {L109} (\bibinfo {year} {2006})}\BibitemShut
  {NoStop}%
\bibitem [{\citenamefont {Alcock}\ \emph {et~al.}(1998)\citenamefont {Alcock}
  \emph {et~al.}}]{AstrophysJLett.499.L9}%
  \BibitemOpen
  \bibfield  {author} {\bibinfo {author} {\bibfnamefont {C.}~\bibnamefont
  {Alcock}} \emph {et~al.},\ }\href
  {http://stacks.iop.org/1538-4357/499/i=1/a=L9} {\bibfield  {journal}
  {\bibinfo  {journal} {Astrophys. J. Lett.}\ }\textbf {\bibinfo {volume}
  {499}},\ \bibinfo {pages} {L9} (\bibinfo {year} {1998})}\BibitemShut
  {NoStop}%
\bibitem [{\citenamefont {Aprile}\ \emph {et~al.}(2018)\citenamefont {Aprile}
  \emph {et~al.}}]{1805.12562}%
  \BibitemOpen
  \bibfield  {author} {\bibinfo {author} {\bibfnamefont {E.}~\bibnamefont
  {Aprile}} \emph {et~al.} (\bibinfo {collaboration} {XENON Collaboration}),\
  }\href {\doibase 10.1103/PhysRevLett.121.111302} {\bibfield  {journal}
  {\bibinfo  {journal} {Phys. Rev. Lett.}\ }\textbf {\bibinfo {volume} {121}},\
  \bibinfo {pages} {111302} (\bibinfo {year} {2018})}\BibitemShut {NoStop}%
\bibitem [{\citenamefont {Akerib}\ \emph {et~al.}(2017)\citenamefont {Akerib}
  \emph {et~al.}}]{PhysRevLett.118.021303}%
  \BibitemOpen
  \bibfield  {author} {\bibinfo {author} {\bibfnamefont {D.~S.}\ \bibnamefont
  {Akerib}} \emph {et~al.} (\bibinfo {collaboration} {LUX Collaboration}),\
  }\href {\doibase 10.1103/PhysRevLett.118.021303} {\bibfield  {journal}
  {\bibinfo  {journal} {Phys. Rev. Lett.}\ }\textbf {\bibinfo {volume} {118}},\
  \bibinfo {pages} {021303} (\bibinfo {year} {2017})}\BibitemShut {NoStop}%
\bibitem [{\citenamefont {Cui}\ \emph {et~al.}(2017)\citenamefont {Cui} \emph
  {et~al.}}]{PhysRevLett.119.181302}%
  \BibitemOpen
  \bibfield  {author} {\bibinfo {author} {\bibfnamefont {X.}~\bibnamefont
  {Cui}} \emph {et~al.} (\bibinfo {collaboration} {PandaX-II Collaboration}),\
  }\href {\doibase 10.1103/PhysRevLett.119.181302} {\bibfield  {journal}
  {\bibinfo  {journal} {Phys. Rev. Lett.}\ }\textbf {\bibinfo {volume} {119}},\
  \bibinfo {pages} {181302} (\bibinfo {year} {2017})}\BibitemShut {NoStop}%
\bibitem [{\citenamefont {Agnese}\ \emph {et~al.}(2018)\citenamefont {Agnese}
  \emph {et~al.}}]{PhysRevD.97.022002}%
  \BibitemOpen
  \bibfield  {author} {\bibinfo {author} {\bibfnamefont {R.}~\bibnamefont
  {Agnese}} \emph {et~al.} (\bibinfo {collaboration} {SuperCDMS
  Collaboration}),\ }\href {\doibase 10.1103/PhysRevD.97.022002} {\bibfield
  {journal} {\bibinfo  {journal} {Phys. Rev. D}\ }\textbf {\bibinfo {volume}
  {97}},\ \bibinfo {pages} {022002} (\bibinfo {year} {2018})}\BibitemShut
  {NoStop}%
\bibitem [{\citenamefont {Angloher}\ \emph {et~al.}(2016)\citenamefont
  {Angloher} \emph {et~al.}}]{EurPhysJC.76.25}%
  \BibitemOpen
  \bibfield  {author} {\bibinfo {author} {\bibfnamefont {G.}~\bibnamefont
  {Angloher}} \emph {et~al.},\ }\href {\doibase 10.1140/epjc/s10052-016-3877-3}
  {\bibfield  {journal} {\bibinfo  {journal} {The Eur. Phys. J. C}\ }\textbf
  {\bibinfo {volume} {76}},\ \bibinfo {pages} {25} (\bibinfo {year}
  {2016})}\BibitemShut {NoStop}%
\bibitem [{\citenamefont {Asztalos}\ \emph {et~al.}(2010)\citenamefont
  {Asztalos}, \citenamefont {Carosi}, \citenamefont {Hagmann}, \citenamefont
  {Kinion}, \citenamefont {van Bibber}, \citenamefont {Hotz}, \citenamefont
  {Rosenberg}, \citenamefont {Rybka}, \citenamefont {Hoskins}, \citenamefont
  {Hwang} \emph {et~al.}}]{PhysRevLett.104.041301}%
  \BibitemOpen
  \bibfield  {author} {\bibinfo {author} {\bibfnamefont {S.~J.}\ \bibnamefont
  {Asztalos}}, \bibinfo {author} {\bibfnamefont {G.}~\bibnamefont {Carosi}},
  \bibinfo {author} {\bibfnamefont {C.}~\bibnamefont {Hagmann}}, \bibinfo
  {author} {\bibfnamefont {D.}~\bibnamefont {Kinion}}, \bibinfo {author}
  {\bibfnamefont {K.}~\bibnamefont {van Bibber}}, \bibinfo {author}
  {\bibfnamefont {M.}~\bibnamefont {Hotz}}, \bibinfo {author} {\bibfnamefont
  {L.~J.}\ \bibnamefont {Rosenberg}}, \bibinfo {author} {\bibfnamefont
  {G.}~\bibnamefont {Rybka}}, \bibinfo {author} {\bibfnamefont
  {J.}~\bibnamefont {Hoskins}}, \bibinfo {author} {\bibfnamefont
  {J.}~\bibnamefont {Hwang}},  \emph {et~al.},\ }\href {\doibase
  10.1103/PhysRevLett.104.041301} {\bibfield  {journal} {\bibinfo  {journal}
  {Phys. Rev. Lett.}\ }\textbf {\bibinfo {volume} {104}},\ \bibinfo {pages}
  {041301} (\bibinfo {year} {2010})}\BibitemShut {NoStop}%
\bibitem [{\citenamefont {Arik}\ \emph {et~al.}(2014)\citenamefont {Arik} \emph
  {et~al.}}]{PhysRevLett.112.091302}%
  \BibitemOpen
  \bibfield  {author} {\bibinfo {author} {\bibfnamefont {M.}~\bibnamefont
  {Arik}} \emph {et~al.} (\bibinfo {collaboration} {CAST Collaboration}),\
  }\href {\doibase 10.1103/PhysRevLett.112.091302} {\bibfield  {journal}
  {\bibinfo  {journal} {Phys. Rev. Lett.}\ }\textbf {\bibinfo {volume} {112}},\
  \bibinfo {pages} {091302} (\bibinfo {year} {2014})}\BibitemShut {NoStop}%
\bibitem [{\citenamefont {Graham}\ \emph {et~al.}(2015)\citenamefont {Graham},
  \citenamefont {Irastorza}, \citenamefont {Lamoreaux}, \citenamefont
  {Lindner},\ and\ \citenamefont {van Bibber}}]{AnnRevNuclPartSci.65.485}%
  \BibitemOpen
  \bibfield  {author} {\bibinfo {author} {\bibfnamefont {P.~W.}\ \bibnamefont
  {Graham}}, \bibinfo {author} {\bibfnamefont {I.~G.}\ \bibnamefont
  {Irastorza}}, \bibinfo {author} {\bibfnamefont {S.~K.}\ \bibnamefont
  {Lamoreaux}}, \bibinfo {author} {\bibfnamefont {A.}~\bibnamefont {Lindner}},
  \ and\ \bibinfo {author} {\bibfnamefont {K.~A.}\ \bibnamefont {van Bibber}},\
  }\href {\doibase 10.1146/annurev-nucl-102014-022120} {\bibfield  {journal}
  {\bibinfo  {journal} {Annu. Rev. Nucl. Part. Sci.}\ }\textbf {\bibinfo
  {volume} {65}},\ \bibinfo {pages} {485} (\bibinfo {year} {2015})}\BibitemShut
  {NoStop}%
\bibitem [{\citenamefont {Carr}\ \emph {et~al.}(2016)\citenamefont {Carr},
  \citenamefont {K\"uhnel},\ and\ \citenamefont
  {Sandstad}}]{PhysRevD.94.083504}%
  \BibitemOpen
  \bibfield  {author} {\bibinfo {author} {\bibfnamefont {B.}~\bibnamefont
  {Carr}}, \bibinfo {author} {\bibfnamefont {F.}~\bibnamefont {K\"uhnel}}, \
  and\ \bibinfo {author} {\bibfnamefont {M.}~\bibnamefont {Sandstad}},\ }\href
  {\doibase 10.1103/PhysRevD.94.083504} {\bibfield  {journal} {\bibinfo
  {journal} {Phys. Rev. D}\ }\textbf {\bibinfo {volume} {94}},\ \bibinfo
  {pages} {083504} (\bibinfo {year} {2016})}\BibitemShut {NoStop}%
\bibitem [{\citenamefont {Landau}\ and\ \citenamefont
  {Lifshitz}(1976)}]{LandauMechanics}%
  \BibitemOpen
  \bibfield  {author} {\bibinfo {author} {\bibfnamefont {L.~D.}\ \bibnamefont
  {Landau}}\ and\ \bibinfo {author} {\bibfnamefont {E.~M.}\ \bibnamefont
  {Lifshitz}},\ }\href@noop {} {\emph {\bibinfo {title} {Mechanics Third
  Edition}}}\ (\bibinfo  {publisher} {Elsevier},\ \bibinfo {year}
  {1976})\BibitemShut {NoStop}%
\bibitem [{\citenamefont {Hoang}\ \emph {et~al.}(2016)\citenamefont {Hoang},
  \citenamefont {Ma}, \citenamefont {Ahn}, \citenamefont {Bang}, \citenamefont
  {Robicheaux}, \citenamefont {Yin},\ and\ \citenamefont
  {Li}}]{PhysRevLett.117.123604}%
  \BibitemOpen
  \bibfield  {author} {\bibinfo {author} {\bibfnamefont {T.~M.}\ \bibnamefont
  {Hoang}}, \bibinfo {author} {\bibfnamefont {Y.}~\bibnamefont {Ma}}, \bibinfo
  {author} {\bibfnamefont {J.}~\bibnamefont {Ahn}}, \bibinfo {author}
  {\bibfnamefont {J.}~\bibnamefont {Bang}}, \bibinfo {author} {\bibfnamefont
  {F.}~\bibnamefont {Robicheaux}}, \bibinfo {author} {\bibfnamefont {Z.-Q.}\
  \bibnamefont {Yin}}, \ and\ \bibinfo {author} {\bibfnamefont
  {T.}~\bibnamefont {Li}},\ }\href {\doibase 10.1103/PhysRevLett.117.123604}
  {\bibfield  {journal} {\bibinfo  {journal} {Phys. Rev. Lett.}\ }\textbf
  {\bibinfo {volume} {117}},\ \bibinfo {pages} {123604} (\bibinfo {year}
  {2016})}\BibitemShut {NoStop}%
\bibitem [{\citenamefont {Rider}\ \emph {et~al.}(2018)\citenamefont {Rider},
  \citenamefont {Blakemore}, \citenamefont {Gratta},\ and\ \citenamefont
  {Moore}}]{PhysRevA.97.013842}%
  \BibitemOpen
  \bibfield  {author} {\bibinfo {author} {\bibfnamefont {A.~D.}\ \bibnamefont
  {Rider}}, \bibinfo {author} {\bibfnamefont {C.~P.}\ \bibnamefont
  {Blakemore}}, \bibinfo {author} {\bibfnamefont {G.}~\bibnamefont {Gratta}}, \
  and\ \bibinfo {author} {\bibfnamefont {D.~C.}\ \bibnamefont {Moore}},\ }\href
  {\doibase 10.1103/PhysRevA.97.013842} {\bibfield  {journal} {\bibinfo
  {journal} {Phys. Rev. A}\ }\textbf {\bibinfo {volume} {97}},\ \bibinfo
  {pages} {013842} (\bibinfo {year} {2018})}\BibitemShut {NoStop}%
\bibitem [{\citenamefont {Monteiro}\ \emph {et~al.}(2017)\citenamefont
  {Monteiro}, \citenamefont {Ghosh}, \citenamefont {Fine},\ and\ \citenamefont
  {Moore}}]{PhysRevA.96.063841}%
  \BibitemOpen
  \bibfield  {author} {\bibinfo {author} {\bibfnamefont {F.}~\bibnamefont
  {Monteiro}}, \bibinfo {author} {\bibfnamefont {S.}~\bibnamefont {Ghosh}},
  \bibinfo {author} {\bibfnamefont {A.~G.}\ \bibnamefont {Fine}}, \ and\
  \bibinfo {author} {\bibfnamefont {D.~C.}\ \bibnamefont {Moore}},\ }\href
  {\doibase 10.1103/PhysRevA.96.063841} {\bibfield  {journal} {\bibinfo
  {journal} {Phys. Rev. A}\ }\textbf {\bibinfo {volume} {96}},\ \bibinfo
  {pages} {063841} (\bibinfo {year} {2017})}\BibitemShut {NoStop}%
\bibitem [{\citenamefont {Li}\ \emph {et~al.}(2011)\citenamefont {Li},
  \citenamefont {Kheifets},\ and\ \citenamefont {Raizen}}]{NatPhys.7.527}%
  \BibitemOpen
  \bibfield  {author} {\bibinfo {author} {\bibfnamefont {T.}~\bibnamefont
  {Li}}, \bibinfo {author} {\bibfnamefont {S.}~\bibnamefont {Kheifets}}, \ and\
  \bibinfo {author} {\bibfnamefont {M.~G.}\ \bibnamefont {Raizen}},\ }\href
  {http://dx.doi.org/10.1038/nphys1952} {\bibfield  {journal} {\bibinfo
  {journal} {Nat. Phys.}\ }\textbf {\bibinfo {volume} {7}},\ \bibinfo {pages}
  {527} (\bibinfo {year} {2011})}\BibitemShut {NoStop}%
\bibitem [{\citenamefont {Ashkin}\ and\ \citenamefont
  {Dziedzic}(1971)}]{ApplPhysLett.19.283}%
  \BibitemOpen
  \bibfield  {author} {\bibinfo {author} {\bibfnamefont {A.}~\bibnamefont
  {Ashkin}}\ and\ \bibinfo {author} {\bibfnamefont {J.~M.}\ \bibnamefont
  {Dziedzic}},\ }\href {\doibase 10.1063/1.1653919} {\bibfield  {journal}
  {\bibinfo  {journal} {Appl. Phys. Lett.}\ }\textbf {\bibinfo {volume} {19}},\
  \bibinfo {pages} {283} (\bibinfo {year} {1971})}\BibitemShut {NoStop}%
\bibitem [{\citenamefont {Gieseler}\ \emph {et~al.}(2012)\citenamefont
  {Gieseler}, \citenamefont {Deutsch}, \citenamefont {Quidant},\ and\
  \citenamefont {Novotny}}]{PhysRevLett.109.103603}%
  \BibitemOpen
  \bibfield  {author} {\bibinfo {author} {\bibfnamefont {J.}~\bibnamefont
  {Gieseler}}, \bibinfo {author} {\bibfnamefont {B.}~\bibnamefont {Deutsch}},
  \bibinfo {author} {\bibfnamefont {R.}~\bibnamefont {Quidant}}, \ and\
  \bibinfo {author} {\bibfnamefont {L.}~\bibnamefont {Novotny}},\ }\href
  {\doibase 10.1103/PhysRevLett.109.103603} {\bibfield  {journal} {\bibinfo
  {journal} {Phys. Rev. Lett.}\ }\textbf {\bibinfo {volume} {109}},\ \bibinfo
  {pages} {103603} (\bibinfo {year} {2012})}\BibitemShut {NoStop}%
\bibitem [{\citenamefont {Asenbaum}\ \emph {et~al.}(2013)\citenamefont
  {Asenbaum}, \citenamefont {Kuhn}, \citenamefont {Nimmrichter}, \citenamefont
  {Sezer},\ and\ \citenamefont {Arndt}}]{NatCommun.4.2743}%
  \BibitemOpen
  \bibfield  {author} {\bibinfo {author} {\bibfnamefont {P.}~\bibnamefont
  {Asenbaum}}, \bibinfo {author} {\bibfnamefont {S.}~\bibnamefont {Kuhn}},
  \bibinfo {author} {\bibfnamefont {S.}~\bibnamefont {Nimmrichter}}, \bibinfo
  {author} {\bibfnamefont {U.}~\bibnamefont {Sezer}}, \ and\ \bibinfo {author}
  {\bibfnamefont {M.}~\bibnamefont {Arndt}},\ }\href
  {http://dx.doi.org/10.1038/ncomms3743} {\bibfield  {journal} {\bibinfo
  {journal} {Nat. Commun.}\ }\textbf {\bibinfo {volume} {4}},\ \bibinfo {pages}
  {2743} (\bibinfo {year} {2013})}\BibitemShut {NoStop}%
\bibitem [{\citenamefont {Mazilu}\ \emph {et~al.}(2016)\citenamefont {Mazilu},
  \citenamefont {Arita}, \citenamefont {Vettenburg}, \citenamefont {Au\~n\'on},
  \citenamefont {Wright},\ and\ \citenamefont {Dholakia}}]{PhysRevA.94.053821}%
  \BibitemOpen
  \bibfield  {author} {\bibinfo {author} {\bibfnamefont {M.}~\bibnamefont
  {Mazilu}}, \bibinfo {author} {\bibfnamefont {Y.}~\bibnamefont {Arita}},
  \bibinfo {author} {\bibfnamefont {T.}~\bibnamefont {Vettenburg}}, \bibinfo
  {author} {\bibfnamefont {J.~M.}\ \bibnamefont {Au\~n\'on}}, \bibinfo {author}
  {\bibfnamefont {E.~M.}\ \bibnamefont {Wright}}, \ and\ \bibinfo {author}
  {\bibfnamefont {K.}~\bibnamefont {Dholakia}},\ }\href {\doibase
  10.1103/PhysRevA.94.053821} {\bibfield  {journal} {\bibinfo  {journal} {Phys.
  Rev. A}\ }\textbf {\bibinfo {volume} {94}},\ \bibinfo {pages} {053821}
  (\bibinfo {year} {2016})}\BibitemShut {NoStop}%
\bibitem [{\citenamefont {Ranjit}\ \emph {et~al.}(2016)\citenamefont {Ranjit},
  \citenamefont {Cunningham}, \citenamefont {Casey},\ and\ \citenamefont
  {Geraci}}]{PhysRevA.93.053801}%
  \BibitemOpen
  \bibfield  {author} {\bibinfo {author} {\bibfnamefont {G.}~\bibnamefont
  {Ranjit}}, \bibinfo {author} {\bibfnamefont {M.}~\bibnamefont {Cunningham}},
  \bibinfo {author} {\bibfnamefont {K.}~\bibnamefont {Casey}}, \ and\ \bibinfo
  {author} {\bibfnamefont {A.~A.}\ \bibnamefont {Geraci}},\ }\href {\doibase
  10.1103/PhysRevA.93.053801} {\bibfield  {journal} {\bibinfo  {journal} {Phys.
  Rev. A}\ }\textbf {\bibinfo {volume} {93}},\ \bibinfo {pages} {053801}
  (\bibinfo {year} {2016})}\BibitemShut {NoStop}%
\bibitem [{\citenamefont {Fonseca}\ \emph {et~al.}(2016)\citenamefont
  {Fonseca}, \citenamefont {Aranas}, \citenamefont {Millen}, \citenamefont
  {Monteiro},\ and\ \citenamefont {Barker}}]{PhysRevLett.117.173602}%
  \BibitemOpen
  \bibfield  {author} {\bibinfo {author} {\bibfnamefont {P.~Z.~G.}\
  \bibnamefont {Fonseca}}, \bibinfo {author} {\bibfnamefont {E.~B.}\
  \bibnamefont {Aranas}}, \bibinfo {author} {\bibfnamefont {J.}~\bibnamefont
  {Millen}}, \bibinfo {author} {\bibfnamefont {T.~S.}\ \bibnamefont
  {Monteiro}}, \ and\ \bibinfo {author} {\bibfnamefont {P.~F.}\ \bibnamefont
  {Barker}},\ }\href {\doibase 10.1103/PhysRevLett.117.173602} {\bibfield
  {journal} {\bibinfo  {journal} {Phys. Rev. Lett.}\ }\textbf {\bibinfo
  {volume} {117}},\ \bibinfo {pages} {173602} (\bibinfo {year}
  {2016})}\BibitemShut {NoStop}%
\bibitem [{\citenamefont {Vovrosh}\ \emph {et~al.}(2017)\citenamefont
  {Vovrosh}, \citenamefont {Rashid}, \citenamefont {Hempston}, \citenamefont
  {Bateman}, \citenamefont {Paternostro},\ and\ \citenamefont
  {Ulbricht}}]{JOptSocAmB.34.1421}%
  \BibitemOpen
  \bibfield  {author} {\bibinfo {author} {\bibfnamefont {J.}~\bibnamefont
  {Vovrosh}}, \bibinfo {author} {\bibfnamefont {M.}~\bibnamefont {Rashid}},
  \bibinfo {author} {\bibfnamefont {D.}~\bibnamefont {Hempston}}, \bibinfo
  {author} {\bibfnamefont {J.}~\bibnamefont {Bateman}}, \bibinfo {author}
  {\bibfnamefont {M.}~\bibnamefont {Paternostro}}, \ and\ \bibinfo {author}
  {\bibfnamefont {H.}~\bibnamefont {Ulbricht}},\ }\href {\doibase
  10.1364/JOSAB.34.001421} {\bibfield  {journal} {\bibinfo  {journal} {J. Opt.
  Soc. Am. B}\ }\textbf {\bibinfo {volume} {34}},\ \bibinfo {pages} {1421}
  (\bibinfo {year} {2017})}\BibitemShut {NoStop}%
\bibitem [{\citenamefont {Kiesel}\ \emph {et~al.}(2013)\citenamefont {Kiesel},
  \citenamefont {Blaser}, \citenamefont {Deli{\'c}}, \citenamefont {Grass},
  \citenamefont {Kaltenbaek},\ and\ \citenamefont
  {Aspelmeyer}}]{ProcNatAcadSci.35.14180}%
  \BibitemOpen
  \bibfield  {author} {\bibinfo {author} {\bibfnamefont {N.}~\bibnamefont
  {Kiesel}}, \bibinfo {author} {\bibfnamefont {F.}~\bibnamefont {Blaser}},
  \bibinfo {author} {\bibfnamefont {U.}~\bibnamefont {Deli{\'c}}}, \bibinfo
  {author} {\bibfnamefont {D.}~\bibnamefont {Grass}}, \bibinfo {author}
  {\bibfnamefont {R.}~\bibnamefont {Kaltenbaek}}, \ and\ \bibinfo {author}
  {\bibfnamefont {M.}~\bibnamefont {Aspelmeyer}},\ }\href {\doibase
  10.1073/pnas.1309167110} {\bibfield  {journal} {\bibinfo  {journal} {Proc.
  Natl. Acad. Sci.}\ }\textbf {\bibinfo {volume} {110}},\ \bibinfo {pages}
  {14180} (\bibinfo {year} {2013})}\BibitemShut {NoStop}%
\bibitem [{\citenamefont {Martynov}\ \emph {et~al.}(2016)\citenamefont
  {Martynov} \emph {et~al.}}]{PhysRevD.93.112004}%
  \BibitemOpen
  \bibfield  {author} {\bibinfo {author} {\bibfnamefont {D.~V.}\ \bibnamefont
  {Martynov}} \emph {et~al.},\ }\href {\doibase 10.1103/PhysRevD.93.112004}
  {\bibfield  {journal} {\bibinfo  {journal} {Phys. Rev. D}\ }\textbf {\bibinfo
  {volume} {93}},\ \bibinfo {pages} {112004} (\bibinfo {year}
  {2016})}\BibitemShut {NoStop}%
\bibitem [{\citenamefont {Acernese}\ \emph {et~al.}(2015)\citenamefont
  {Acernese} \emph {et~al.}}]{ClassQuantGrav.32.024001}%
  \BibitemOpen
  \bibfield  {author} {\bibinfo {author} {\bibfnamefont {F.}~\bibnamefont
  {Acernese}} \emph {et~al.},\ }\href
  {http://stacks.iop.org/0264-9381/32/i=2/a=024001} {\bibfield  {journal}
  {\bibinfo  {journal} {Class. Quantum Grav.}\ }\textbf {\bibinfo {volume}
  {32}},\ \bibinfo {pages} {024001} (\bibinfo {year} {2015})}\BibitemShut
  {NoStop}%
\bibitem [{\citenamefont {Somiya}(2012)}]{ClassQuantGrav.29.124007}%
  \BibitemOpen
  \bibfield  {author} {\bibinfo {author} {\bibfnamefont {K.}~\bibnamefont
  {Somiya}},\ }\href {http://stacks.iop.org/0264-9381/29/i=12/a=124007}
  {\bibfield  {journal} {\bibinfo  {journal} {Class. Quantum Grav.}\ }\textbf
  {\bibinfo {volume} {29}},\ \bibinfo {pages} {124007} (\bibinfo {year}
  {2012})}\BibitemShut {NoStop}%
\bibitem [{\citenamefont {Herzog-Arbeitman}\ \emph {et~al.}(2018)\citenamefont
  {Herzog-Arbeitman}, \citenamefont {Lisanti}, \citenamefont {Madau},\ and\
  \citenamefont {Necib}}]{PhysRevLett.120.041102}%
  \BibitemOpen
  \bibfield  {author} {\bibinfo {author} {\bibfnamefont {J.}~\bibnamefont
  {Herzog-Arbeitman}}, \bibinfo {author} {\bibfnamefont {M.}~\bibnamefont
  {Lisanti}}, \bibinfo {author} {\bibfnamefont {P.}~\bibnamefont {Madau}}, \
  and\ \bibinfo {author} {\bibfnamefont {L.}~\bibnamefont {Necib}},\ }\href
  {\doibase 10.1103/PhysRevLett.120.041102} {\bibfield  {journal} {\bibinfo
  {journal} {Phys. Rev. Lett.}\ }\textbf {\bibinfo {volume} {120}},\ \bibinfo
  {pages} {041102} (\bibinfo {year} {2018})}\BibitemShut {NoStop}%
\bibitem [{\citenamefont {Catena}\ and\ \citenamefont
  {Ullio}(2010)}]{JCosmolAstropartPhys.2010.004}%
  \BibitemOpen
  \bibfield  {author} {\bibinfo {author} {\bibfnamefont {R.}~\bibnamefont
  {Catena}}\ and\ \bibinfo {author} {\bibfnamefont {P.}~\bibnamefont {Ullio}},\
  }\href {http://stacks.iop.org/1475-7516/2010/i=08/a=004} {\bibfield
  {journal} {\bibinfo  {journal} {J. Cosmol. Astropart. Phys.}\ }\textbf
  {\bibinfo {volume} {2010}},\ \bibinfo {pages} {004} (\bibinfo {year}
  {2010})}\BibitemShut {NoStop}%
\bibitem [{\citenamefont {Tanabashi}\ \emph {et~al.}(2018)\citenamefont
  {Tanabashi} \emph {et~al.}}]{PDG}%
  \BibitemOpen
  \bibfield  {author} {\bibinfo {author} {\bibfnamefont {M.}~\bibnamefont
  {Tanabashi}} \emph {et~al.},\ }\href {http://pdg.lbl.gov/} {\bibfield
  {journal} {\bibinfo  {journal} {Phys. Rev. D}\ }\textbf {\bibinfo {volume}
  {98}},\ \bibinfo {pages} {030001} (\bibinfo {year} {2018})}\BibitemShut
  {NoStop}%
\bibitem [{\citenamefont {Aston}\ \emph {et~al.}(2012)\citenamefont {Aston}
  \emph {et~al.}}]{ClassQuantumGrav.29.235004}%
  \BibitemOpen
  \bibfield  {author} {\bibinfo {author} {\bibfnamefont {S.~M.}\ \bibnamefont
  {Aston}} \emph {et~al.},\ }\href
  {http://stacks.iop.org/0264-9381/29/i=23/a=235004} {\bibfield  {journal}
  {\bibinfo  {journal} {Class. Quantum Grav.}\ }\textbf {\bibinfo {volume}
  {29}},\ \bibinfo {pages} {235004} (\bibinfo {year} {2012})}\BibitemShut
  {NoStop}%
\bibitem [{\citenamefont {Punturo}\ \emph {et~al.}(2010)\citenamefont {Punturo}
  \emph {et~al.}}]{ClassQuantGrav.27.194002}%
  \BibitemOpen
  \bibfield  {author} {\bibinfo {author} {\bibfnamefont {M.}~\bibnamefont
  {Punturo}} \emph {et~al.},\ }\href
  {http://stacks.iop.org/0264-9381/27/i=19/a=194002} {\bibfield  {journal}
  {\bibinfo  {journal} {Class. Quantum Grav.}\ }\textbf {\bibinfo {volume}
  {27}},\ \bibinfo {pages} {194002} (\bibinfo {year} {2010})}\BibitemShut
  {NoStop}%
\bibitem [{\citenamefont {Hild}\ \emph {et~al.}(2008)\citenamefont {Hild},
  \citenamefont {Chelkowski},\ and\ \citenamefont {Freise}}]{0810.0604}%
  \BibitemOpen
  \bibfield  {author} {\bibinfo {author} {\bibfnamefont {S.}~\bibnamefont
  {Hild}}, \bibinfo {author} {\bibfnamefont {S.}~\bibnamefont {Chelkowski}}, \
  and\ \bibinfo {author} {\bibfnamefont {A.}~\bibnamefont {Freise}},\ }\href
  {https://arxiv.org/abs/0810.0604v2} {\  (\bibinfo {year} {2008})},\ \Eprint
  {http://arxiv.org/abs/arXiv:0810.0604} {arXiv:0810.0604} \BibitemShut
  {NoStop}%
\bibitem [{\citenamefont {Hild}\ \emph {et~al.}(2010)\citenamefont {Hild},
  \citenamefont {Chelkowski}, \citenamefont {Freise}, \citenamefont {Franc},
  \citenamefont {Morgado}, \citenamefont {Flaminio},\ and\ \citenamefont
  {DeSalvo}}]{ClassQuantGrav.27.015003}%
  \BibitemOpen
  \bibfield  {author} {\bibinfo {author} {\bibfnamefont {S.}~\bibnamefont
  {Hild}}, \bibinfo {author} {\bibfnamefont {S.}~\bibnamefont {Chelkowski}},
  \bibinfo {author} {\bibfnamefont {A.}~\bibnamefont {Freise}}, \bibinfo
  {author} {\bibfnamefont {J.}~\bibnamefont {Franc}}, \bibinfo {author}
  {\bibfnamefont {N.}~\bibnamefont {Morgado}}, \bibinfo {author} {\bibfnamefont
  {R.}~\bibnamefont {Flaminio}}, \ and\ \bibinfo {author} {\bibfnamefont
  {R.}~\bibnamefont {DeSalvo}},\ }\href
  {http://stacks.iop.org/0264-9381/27/i=1/a=015003} {\bibfield  {journal}
  {\bibinfo  {journal} {Classical and Quantum Gravity}\ }\textbf {\bibinfo
  {volume} {27}},\ \bibinfo {pages} {015003} (\bibinfo {year}
  {2010})}\BibitemShut {NoStop}%
\bibitem [{\citenamefont {Hall}\ \emph {et~al.}(2018)\citenamefont {Hall},
  \citenamefont {Adhikari}, \citenamefont {Frolov}, \citenamefont {M\"uller},\
  and\ \citenamefont {Pospelov}}]{PhysRevD.98.083019}%
  \BibitemOpen
  \bibfield  {author} {\bibinfo {author} {\bibfnamefont {E.~D.}\ \bibnamefont
  {Hall}}, \bibinfo {author} {\bibfnamefont {R.~X.}\ \bibnamefont {Adhikari}},
  \bibinfo {author} {\bibfnamefont {V.~V.}\ \bibnamefont {Frolov}}, \bibinfo
  {author} {\bibfnamefont {H.}~\bibnamefont {M\"uller}}, \ and\ \bibinfo
  {author} {\bibfnamefont {M.}~\bibnamefont {Pospelov}},\ }\href {\doibase
  10.1103/PhysRevD.98.083019} {\bibfield  {journal} {\bibinfo  {journal} {Phys.
  Rev. D}\ }\textbf {\bibinfo {volume} {98}},\ \bibinfo {pages} {083019}
  (\bibinfo {year} {2018})}\BibitemShut {NoStop}%
\end{thebibliography}%

\end{document}